\documentclass[aps,prd,nofootinbib]{revtex4}
\usepackage{graphicx}

\begin{document}

\title{Stability of Schwarzschild-$f(R)$ gravity thin-shell wormholes}
\author{Alina Khaybullina}
\email{kalina89-rf@mail.ru}
\affiliation{Zel'dovich International Center for Astrophysics, Bashkir State Pedagogical University, 3A, October Revolution Street, Ufa 450008, RB, Russia}
\author{Gulira Tuleganova}
\email{gulira.tuleganova@yandex.ru}
\affiliation{Zel'dovich International Center for Astrophysics, Bashkir State Pedagogical University, 3A, October Revolution Street, Ufa 450008, RB, Russia}
\date{\today}

\begin{abstract}
Mazharimousavi and Halilsoy [1] recently proposed wormhole solutions in $f(R)$-gravity that satisfy energy conditions but are unstable. We show here that stability could still be achieved for thin-shell wormholes obtained by gluing the wormholes in $f(R)$-gravity with the exterior Schwarzschild vacuum. Using the new geometrical constraints from thin-shell ''mass'' and from external ''force'' developed by Garcia, Lobo and Visser, we demarcate and analyze the stability regions.
\end{abstract}
\maketitle


\section{Introduction}

Wormholes are geometrical handles in spacetime that connect two universes or
two distant regions of spacetime and are solutions of Einstein's general
relativity including other theories of gravity. The subject has received
considerable attention after the influential theoretical work by Morris and
Thorne [2]. While there exist enormous work on black holes, especially
directed to observing the signatures of a supermassive black hole at our
galactic center, relatively less work is available on wormholes. This
scenario has now drastically changed due to the recent discovery that
thin-shell wormholes have the ability to mimic recently observed\
gravitational ring-down post-merger waves that have heretofore been believed
to be characteristic exclusively of black hole horizon [3-6]. Therefore, the
study of stability of thin-shell wormholes is of paramount importance.

Morris-Thorne wormholes within Einstein's theory of general relativity (GR)
require exotic\ ghost matter (i.e., matter that violates the Null Energy
Condition \footnote{%
That is, matter having $\rho +p_{r}<0$, $\rho +p_{t}<0$, where $\rho $ is
the energy density, $p_{r}$ and $p_{t}$ are radial and tranverse components
of pressure.}) for their construction, as shown by Hochberg and Visser [7].
While all experiments to date have failed to directly detect astrophysical
ghost matter, modified gravity theories such as the $f(R)$-gravity (where $R$
is the Ricci scalar) seems to provide a promising avenue to look at
gravitational from a more general perspective beyond Einstein's theory (for
which $f(R)=R$). There exist extensive works on black holes and wormholes in
$f(R)$-gravity. Some key references, though by no means exhaustive, might be
mentioned, see e.g., [8-12]. Excellent reviews on $f(R)$-gravity are also
available [13-16].

On the state-of-the-art of $f(R)$-gravity, a thorough update on the
existence, stability and thermodynamics of different configurations is in
order. The existence problem concerns the conditions necessary for the
existence of configurations in $f(R)$-gravity that are analogous to the ones
in GR. The strategy to obtain those conditions is to employ what is called a
\textit{near-horizon test} in [17,18] exploiting the regularity of GR black
hole horizon. The regularity allows the metric functions to be analytically
expanded around the horizon and when they are put back in the field
equations of $f(R)$-gravity, they yield the necessary conditions on the form
of $f(R)$ admitting analog black holes or $f(R)$-black holes. The remarkable
result obtained in [17,18] is that just any arbitrary \textit{polynomial}
form of $f(R)$ do\textit{\ }not admit analog black holes. A classic example
is that the existence of analog Schwarzschild black hole requires that $f(R)$%
-gravity must be of the specific form $f(R)=\alpha \sqrt{R+\beta }$, where $%
\alpha $,$\beta $ are constants. Analog Reissner-Nordstr\"{o}m black hole
requires an infinite series form of $f(R)$ with the coefficents determined
by the horizon radius [19].

As to the question of stability, a viable $f(R)$-gravity solution should be
stable, ghost-free and should admit Newtonian and Post-Newtonian\ limits
including correct cosmological dynamics. This means that for the stability
of analog black holes, the near-horizon test yielding the form of $f(R)$
should be further constrained by the conditions that $\frac{df}{dR}>0$ and $%
\frac{d^{2}f}{dR^{2}}$ $>0$ [19]. From a different viewpoint, Myung, Moon
and Son [20] converted the $f(R)$-gravity into a Brans-Dicke theory with
massive scalaron and showed that the $f(R)$ analog of Schwarzschild black
hole is stable against the external perturbations if the scalaron mass
squared is not negative (tachyonic mass) that is ensured by the condition $%
\left. \frac{d^{2}f}{dR^{2}}\right\vert _{R=0}>0$. The opposite of this
stability inequality is the Dolgov-Kawasaki\textit{\ }instability condition $%
\frac{d^{2}f}{dR^{2}}$ $<0$ applicable to perturbations in several distinct
spacetimes [21-25]. Instability of analog Kerr black holes in $f(R)$-gravity
has also been studied [26]. The status of Jebsen-Birkoff theorem and its
stability in $f(R)$-gravity has been studied in [27].

Thermodynamics of analog black holes have been studied [19,28-31]. In [28],
using the Hawking temperature and entropy, the exact expression for heat
capacity and the first law of thermodynamics have been found by using the
Misner-Sharp formalism [19]. Similar thermodynamics dealing with analog
Reissner-Nordstr\"{o}m black hole also exists [29]. Another novel method of
studying thermodynamics and phase transitions of analog black holes is to
use geometrothermodynamical methods [30]. Yokokura [31] generalized the
space-time thermodynamics and then from the equation for entropy balance for
nonequilibrium processes found the new entropy production terms in $f(R)$%
-gravity.

Wormhole solutions including the thin-shell variants have also been obtained
in $f(R)$-gravity [1,32]. Evolving wormholes have been recently found by
Bhattacharya and Chakraborty [33]. Wormholes in the Palatini formulation of $%
f(R)$ gravity have been studied in [34]. Necessary conditions for having
wormholes in $f(R)$-gravity have been obtained in [35]. Some thin-shell
wormholes and their stabilities have been studied in [36-38]. The focus in
this paper is on the two traversable analytical wormhole solutions recently
found by Mazharimousavi and Halilsoy (MH) [1]. These wormholes are however
unstable since, as generically argued in [39], there cannot be ghost-free,
stable wormholes in $f(R)$-gravity. This being the case, it would be of
interest to study the possibility of stability of thin-shell wormholes
obtained by gluing the exterior Schwarzschild vacuum with MH wormholes of $%
f(R)$-gravity\footnote{%
As appropriately commented by an anonymous referee, the present Letter
provides a "graceful exit" to the usual wormhole instabilities.}.

In this Letter, our purpose is to use the new constraints developed by
Garcia, Lobo and Visser (GLV) [40] for identifying the stability regions of
the linearly perturbed spherical motion of the thin-shell moving in the bulk
spacetime. For this purpose, we obtain the thin-shell by gluing the relevant
solutions at some suitable \textquotedblleft standard\textquotedblright\
coordinate radius. In other words, the asymptotic masses on one side will be
the mass of the wormhole in $f(R)$-gravity and on the other side the
Schwarzschild mass. Henceforth, we take $G=1,c=1$ unless specifically
restored.
\section{GLV formalism for stability}

This formalism being relatively new, for the benefit of readers, we explain
below in slightly more detail the GLV formalism [40] for stability of the
thin-shell and the attendent new concepts. The formalism is quite generic,
flexible and robust that can be applied to general spherically symmetric
spacetimes in 4-dimensions. The idea is to surgically graft together two
bulk spacetimes in such a way that no event horizon is formed. This surgery
generates a thin shell (the wormhole throat, where all the exotic matter is
concentrated) between the two bulk spacetimes on either side of the throat.
The thin shell will be free to move in the bulk spacetimes allowing a fully
dynamic stability analysis against spherically symmetric perturbations. The
stability will then be dictated by the properties of the exotic matter
residing on the wormhole throat. The novelty of the formalism is that, apart
from a "mass" constraint from the mass of the shell, it introduces an
entirely \textit{new} hitherto unknown geometrical constraint of external
"force". The motion of the shell is driven by a "potential" $V(a)$ appearing
in the equation of motion ${\frac{1}{2}}\dot{a}^{2}+V(a)=0$ and the
stability under small oscillation about a static solution $a=a_{0}$ (initial
gluing radius) will be given by the condition $V^{\prime \prime }(a_{0})\geq
0$. This is, in short, the strategy of GLV formalism.

We note that GLV method is developed within Einstein's GR having second
order equations, wheras $f(R)$-gravity equations are of fourth order. To
justify the use of GLV method gluing the two solutions from two different
theories, we recall the conformal equivalence of $f(R)$-gravity with GR
minimally coupled to a scalar field with a potential. Mathematically, the
conformal transformation converts the fourth-order $f(R)$ equations into two
second-order equations, one for the Einstein frame metric and the other for
a scalar field $\varphi $. The needed junction condition on extrinsic
curvature in the Einstein frame is the same as in GR [41].

Omitting lengthy details, we shall try to reproduce the salient features of
GLV formalism as cogently as possible, especially the steps leading to the
usable stability constraints. The method starts with gluing together two
spherically symmetric spacetimes by "cut and paste" surgery developed by
Visser [42]. GLV take two generic spherically symmetric spacetimes $\mathcal{%
M}_{\pm }$ possessing the metrics in each as

\begin{equation}
ds_{\pm }^{2}=-e^{2\Phi _{\pm }(r_{\pm })}\left[ 1-\frac{b_{\pm }(r_{\pm })}{%
r_{\pm }}\right] c^{2}dt_{\pm }^{2}+\left[ 1-\frac{b_{\pm }(r_{\pm })}{%
r_{\pm }}\right] ^{-1}dr_{\pm }^{2}+r_{\pm }^{2}d\Omega _{\pm }^{2},
\end{equation}%
where $\Omega =d\theta ^{2}+\sin ^{2}{\theta }d\phi ^{2}$. A single manifold
$\mathcal{M}$ is obtained by gluing the two manifolds $\mathcal{M_{+}}$ and $%
\mathcal{M_{-}}$ at $\Sigma $, i.e., at $f(r,\tau )=r-a(\tau )=0$. That is,
the points $r<a(\tau )$ are excised out of $\mathcal{M}$ and the intrinsic
metric on $\Sigma $ is assumed to have a form \
\begin{equation}
ds_{\Sigma }^{2}=-d\tau ^{2}+a(\tau )^{2}\,(d\theta ^{2}+\sin ^{2}{\theta }%
\,d\phi ^{2}).
\end{equation}

The position of the junction surface is given by $x^{\mu }(\tau ,\theta
,\phi )=[t(\tau ),a(\tau ),\theta ,\phi ]$, and the non-trivial extrinsic
curvature components on both sides of the shell are given by
\begin{eqnarray}
K_{\;\;\theta }^{\theta \;\pm } &=&\pm \frac{1}{a}\,\sqrt{1-\frac{b_{\pm }(a)%
}{a}+\dot{a}^{2}}\;, \\
K_{\;\;\tau }^{\tau \;\pm } &=&\pm \,\left\{ \frac{\ddot{a}+\frac{b_{\pm
}(a)-b_{\pm }^{\prime }(a)a}{2a^{2}}}{\sqrt{1-\frac{b_{\pm }(a)}{a}+\dot{a}%
^{2}}}+\Phi _{\pm }^{\prime }(a)\sqrt{1-\frac{b_{\pm }(a)}{a}+\dot{a}^{2}}%
\right\} \,,
\end{eqnarray}%
where the prime now denotes a derivative with respect to the coordinate $a$
and overdot denotes derivative with respect to $\tau $. Note that $K_{ij}$
is not continuous across $\Sigma $, so one defines $\kappa
_{ij}=K_{ij}^{+}-K_{ij}^{-}$ that satisfy Lanczos equations\ on the shell as
\begin{equation}
S_{\;j}^{i}=-\frac{1}{8\pi }\,(\kappa _{\;j}^{i}-\delta _{\;j}^{i}\kappa
_{\;k}^{k})\,,
\end{equation}%
where $S_{\;j}^{i}$ is the surface stress-energy tensor on $\Sigma $ defined
by $S_{\;j}^{i}=\mathrm{diag}(-\sigma ,\mathcal{P},\mathcal{P})$, $\sigma $
is the surface energy density and $\mathcal{P}$ is the surface pressure
\begin{eqnarray}
\sigma &=&-\frac{1}{4\pi a}\left[ \sqrt{1-\frac{b_{+}(a)}{a}+\dot{a}^{2}}+%
\sqrt{1-\frac{b_{-}(a)}{a}+\dot{a}^{2}}\,\right] , \\
\mathcal{P} &=&\frac{1}{8\pi a}\left[ \frac{1+\dot{a}^{2}+a\ddot{a}-\frac{%
b_{+}(a)+ab_{+}^{\prime }(a)}{2a}}{\sqrt{1-\frac{b_{+}(a)}{a}+\dot{a}^{2}}}+%
\sqrt{1-\frac{b_{+}(a)}{a}+\dot{a}^{2}}\;a\Phi _{+}^{\prime }(a)\right.
\nonumber \\
&&\qquad \left. +\frac{1+\dot{a}^{2}+a\ddot{a}-\frac{b_{-}(a)+ab_{-}^{\prime
}(a)}{2a}}{\sqrt{1-\frac{b_{-}(a)}{a}+\dot{a}^{2}}}+\sqrt{1-\frac{b_{-}(a)}{a%
}+\dot{a}^{2}}\;a\Phi _{-}^{\prime }(a)\right] .
\end{eqnarray}%
\textbf{T}he surface stress $S_{\;j}^{i}$ satisfies a conservation identity
via Lanczos and Gauss-Codazzi equation as
\begin{equation}
S_{\;j|i}^{i}=\left[ T_{\mu \nu }\;e_{\;(j)}^{\mu }n^{\nu }\right]
_{-}^{+}\,,
\end{equation}%
where $n^{\nu }$ is the unite normal to $\Sigma $ and $e_{\;(j)}^{\mu }$ are
the orthonormal basis vectors such that $g_{ij}=g_{\mu \nu }e_{(i)}^{\mu
}e_{(j)}^{\nu }|_{\pm }$. Defining a new term

\begin{equation}
\Xi =\frac{1}{4\pi a}\,\left[ \Phi _{+}^{\prime }(a)\sqrt{1-\frac{b_{+}(a)}{a%
}+\dot{a}^{2}}+\Phi _{-}^{\prime }(a)\sqrt{1-\frac{b_{-}(a)}{a}+\dot{a}^{2}}%
\,\right] \,,
\end{equation}%
the conservation identity (8) can be written as
\begin{equation}
\frac{d(\sigma A)}{d\tau }+\mathcal{P}\,\frac{dA}{d\tau }=\Xi \,A\,\dot{a}\,,
\end{equation}%
where $\dot{a}=\frac{da}{d\tau }$, $A=4\pi a^{2}$ is the surface area of the
shell. The first term represents change in the internal energy of the shell,
while the second term represents the work done by the shell's internal
force, and the third term represents the work done by the external forces
(genesis of the "force" constraint). Assuming the existence of a suitable
function $\sigma (a)$, the conservation identity can be rewritten as
\begin{equation}
\sigma ^{\prime }=-\frac{2}{a}\,(\sigma +\mathcal{P})+\Xi \,,
\end{equation}%
where $\sigma ^{\prime }=d\sigma /da$. The right hand side is the net
discontinuity of the bulk momentum flux and is physically interpreted as the
work done by external "forces" on the thin shell occurring due to $\Phi
_{\pm }^{\prime }\neq 0$. We can rearrange Eq.(6) into the form
\begin{equation}
\sqrt{1-\frac{b_{+}(a)}{a}+\dot{a}^{2}}=-\sqrt{1-\frac{b_{-}(a)}{a}+\dot{a}%
^{2}}-4\pi a\,\sigma (a)\,,
\end{equation}%
which yields the thin-shell equation of motion given by
\begin{equation}
{\frac{1}{2}}\dot{a}^{2}+V(a)=0\,,
\end{equation}%
where the potential $V(a)$ is given by
\begin{eqnarray}
V(a) &=&{\frac{1}{2}}\left\{ 1-{\frac{\bar{b}(a)}{a}}-\left[ \frac{m_{s}(a)}{%
2a}\right] ^{2}-\left[ \frac{\Delta (a)}{m_{s}(a)}\right] ^{2}\right\} \,, \\
\bar{b}(a) &=&\frac{b_{+}(a)+b_{-}(a)}{2},\Delta (a)=\frac{b_{+}(a)-b_{-}(a)%
}{2}.  \nonumber
\end{eqnarray}%
The potential $V(a)$ is seen as a function of the thin-shell mass $%
m_{s}(a)=4\pi a^{2}\,\sigma (a)$ and is the key for stability analysis. It
follows from Eq.(14) that
\begin{equation}
m_{s}(a)=-a\left[ \sqrt{1-{\frac{b_{+}(a)}{a}}-2V(a)}+\sqrt{1-{\frac{b_{-}(a)%
}{a}}-2V(a)}\right] ,
\end{equation}%
the negative sign is required for compatibility with the Lanczos equations
(5). Assume a static solution (at $a_{0}$) to the equation of motion ${\frac{%
1}{2}}\dot{a}^{2}+V(a)=0$, then a Taylor expansion of $V(a)$ around $a_{0}$
to second order yields
\begin{equation}
V(a)=V(a_{0})+V^{\prime }(a_{0})(a-a_{0})+\frac{1}{2}V^{\prime \prime
}(a_{0})(a-a_{0})^{2}+O[(a-a_{0})^{3}]\,.
\end{equation}%
But since we are expanding around a static solution, $\dot{a}_{0}=\ddot{a}%
_{0}=0$, we automatically have $V(a_{0})=V^{\prime }(a_{0})=0$, so it is
sufficient to consider
\begin{equation}
V(a)=\frac{1}{2}V^{\prime \prime }(a_{0})(a-a_{0})^{2}+O[(a-a_{0})^{3}]\,.
\end{equation}%
The static solution at $a_{0}$ is stable if and only if $V^{\prime \prime
}(a_{0})\geq 0$. Since $V(a_{0})=0$, it follows from Eq.(15) that
\begin{equation}
m_{s}(a_{0})=-a_{0}\left\{ \sqrt{1-{\frac{b_{+}(a_{0})}{a_{0}}}}+\sqrt{1-{%
\frac{b_{-}(a_{0})}{a_{0}}}}\right\} .
\end{equation}%
Linearized stability using the condition $V^{\prime \prime }(a_{0})\geq 0$
leads to an inequality on $m_{s}^{\prime \prime }(a)|_{a_{0}}$, that is
called the "mass" constraint given by%
\begin{eqnarray}
m_{s}^{\prime \prime }(a)|_{a_{0}} &\geq &\frac{1}{4a_{0}^{3}}\left[ \frac{%
[b_{+}(a_{0})-a_{0}b_{+}^{\prime }(a_{0})]^{2}}{[1-b_{+}(a_{0})/a_{0}]^{3/2}}%
+\frac{[b_{-}(a_{0})-a_{0}b_{-}^{\prime }(a_{0})]^{2}}{%
[1-b_{-}(a_{0})/a_{0}]^{3/2}}\right]  \nonumber \\
&&+\frac{1}{2}\left[ \frac{b_{+}^{\prime \prime }(a_{0})}{\sqrt{%
1-b_{+}(a_{0})/a_{0}}}+\frac{b_{-}^{\prime \prime }(a_{0})}{\sqrt{%
1-b_{-}(a_{0})/a_{0}}}\right] .
\end{eqnarray}%
Assuming integrability of Eq.(11), lengthy calculations yield the parametric
solution: $\sigma \equiv \sigma (a)$, $\mathcal{P}\equiv \mathcal{P}(a)$ and
$\Xi \equiv \Xi (a)$. The same linearized stability analysis with $V^{\prime
\prime }(a_{0})\geq 0$ also yields inequalities on $\sigma ^{\prime }$, $%
\sigma ^{\prime \prime }$, $\mathcal{P}^{\prime }$,$\mathcal{P}^{\prime
\prime }$, $\Xi ^{\prime }$, $\Xi ^{\prime \prime }$ but the last one is the
most relevant. When $\Phi _{\pm }\neq 0$, there will appear an additional
constraint analogous to the one on $m_{s}^{\prime \prime }(a)|_{a_{0}}$,
called the "force" constraint given by
\begin{eqnarray}
\left. \lbrack 4\pi \,\Xi (a)\,a]^{\prime \prime }\right\vert _{a_{0}} &\leq
&\left. \left\{ \Phi _{+}^{\prime \prime \prime }(a)\sqrt{1-b_{+}(a)/a}+\Phi
_{-}^{\prime \prime \prime }(a)\sqrt{1-b_{-}(a)/a}\right\} \right\vert
_{a_{0}}  \nonumber \\
&&-\left. \left\{ \Phi _{+}^{\prime \prime }(a){\frac{(b_{+}(a)/a)^{\prime }%
}{\sqrt{1-b_{+}(a)/a}}}+\Phi _{-}^{\prime \prime }(a){\frac{%
(b_{-}(a)/a)^{\prime }}{\sqrt{1-b_{-}(a)/a}}}\right\} \right\vert _{a_{0}}
\nonumber \\
&&-{\frac{1}{4}}\left. \left\{ \Phi _{+}^{\prime }(a){\frac{%
[(b_{+}(a)/a)^{\prime 2}}{[1-b_{+}(a)/a]^{3/2}}}+\Phi _{-}^{\prime }(a){%
\frac{[(b_{-}(a)/a)^{\prime 2}}{[1-b_{-}(a)/a]^{3/2}}}\right\} \right\vert
_{a_{0}}  \nonumber \\
&&-{\frac{1}{2}}\left. \left\{ \Phi _{+}^{\prime }(a){\frac{%
(b_{+}(a)/a)^{\prime \prime }}{\sqrt{1-b_{+}(a)/a}}}+\Phi _{-}^{\prime }(a){%
\frac{(b_{-}(a)/a)^{\prime \prime }}{\sqrt{1-b_{-}(a)/a}}}\right\}
\right\vert _{a_{0}}.
\end{eqnarray}%
The same force constraint appears for $\Phi _{\pm }^{\prime }(a_{0})\leq 0$
as well but with only the sign in the inequality \textit{reversed}. The
three inequalities, (19),(20) and the reversed one, are the GLV master
inequalities for deciding stability zones of thin-shell motion.

\section{Stability of thin-shell from Schwarzschild black hole-$f(R)$ wormhole gluing}

\textit{(a) Schwarzschild black hole-MH1 wormhole }

We glue at a common radius $a_{0}$ above the Schwarzschild horizon $r_{\text{%
hor}}=2M$ and wormhole throat $r_{0}$, i.e., at a certain radius $%
r=a_{0}>2M,r_{0}$. The interior regions $r\leq 2M$, $r_{0}$ are surgically
excised out of the respective spacetimes because we don't want the presence
of any horizon in the resultant wormhole. Thus linear spherical
perturbations will be assumed to take place about the radius $a_{0}$.
Casting the Schwarzschild metric into the GLV form of Eq.(1), one obtains

\begin{equation}
b_{+}=2M,\qquad \Phi _{+}=0,
\end{equation}%
and similarly casting the first metric of MH1 wormhole [1] of $f(R)$%
-gravity, viz.,

\begin{equation}
d\tau ^{2}=-\left( 1+\frac{r_{0}}{r}\right) ^{-2}dt^{2}+\frac{dr^{2}}{1-%
\frac{r_{0}^{2}}{r^{2}}}+r^{2}(d\theta ^{2}+\sin ^{2}{\theta }d\phi ^{2}),
\end{equation}%
in the form of GLV metric, we get

\begin{equation}
b_{-}=\frac{r_{0}^{2}}{r},~~\Phi _{-}=-\frac{1}{2}\ln \left[ \left( 1-\frac{%
r_{0}}{r}\right) \left( 1+\frac{r_{0}}{r}\right) ^{3}\right] .
\end{equation}

The redshift function $\Phi _{-}(r)$, now denoted without subscript simply
as $\Phi (r)$, happens to be a logarithmic function showing divergence at $%
r=r_{0}$ but from the physical standpoint this does not pose any problem
since we already made the choice that the accessible region has $r>r_{0}$ -
the throat radius of the wormhole.

Stability zone of thin shell depends on the sign of $\Phi ^{\prime }(a_{0})$%
, which for the MH1 solution is given by%
\begin{equation}
\Phi ^{\prime }(a_{0})=\frac{\left( a_{0}-2r_{0}\right) r_{0}}{\left(
a_{0}^{2}-r_{0}^{2}\right) a_{0}}.
\end{equation}%
There occurs two cases. Case 1: $r=a_{0}\in (r_{0},2r_{0}]$, for which we
get $\Phi ^{\prime }(a_{0})\leq 0$ and Case 2: $r=a_{0}\in \lbrack
2r_{0},\infty )$, for which we get $\Phi ^{\prime }(a_{0})\geq 0$. In either
case, there will occur the effect of external "force"\ influencing the thin
shell motion. Introducing the above functions in the inequalities (19,20),
and defining the dimensionless variables

\begin{equation}
x=\frac{M}{a_{0}},\text{ }y=\frac{r_{0}}{M},
\end{equation}%
we find, respectively

\begin{equation}
m_{s}^{\prime \prime }(a)|_{a_{0}}\geq f(x,y)=x\left( \frac{2}{\sqrt{1-2x}}+%
\frac{x}{(1-2x)^{3/2}}-\frac{xy^{2}(2x^{2}y^{2}-3)}{(1-x^{2}y^{2})^{3/2}}%
\right) ,
\end{equation}

\begin{eqnarray}
\lbrack 4\pi \Xi (a)a]^{\prime \prime }|_{a_{0}} &\geq &g_{1}(x,y)=\frac{%
xy(6-24xy-5x^{2}y^{2}+30x^{3}y^{3}+2x^{4}y^{4}-12x^{5}y^{5})}{%
(1-x^{2}y^{2})^{5/2}}\text{, } \\
&&\left( \Phi _{\pm }^{\prime }(a_{0})\leq 0,\text{ Case 1}\right) \\
\lbrack 4\pi \Xi (a)a]^{\prime \prime }|_{a_{0}} &\leq &g_{1}(x,y).\text{ \
\ }\left( \Phi _{\pm }^{\prime }(a_{0})\geq 0,\text{ Case 2}\right)
\end{eqnarray}

\begin{figure}[!ht]
  \centerline{\includegraphics[scale=0.5]{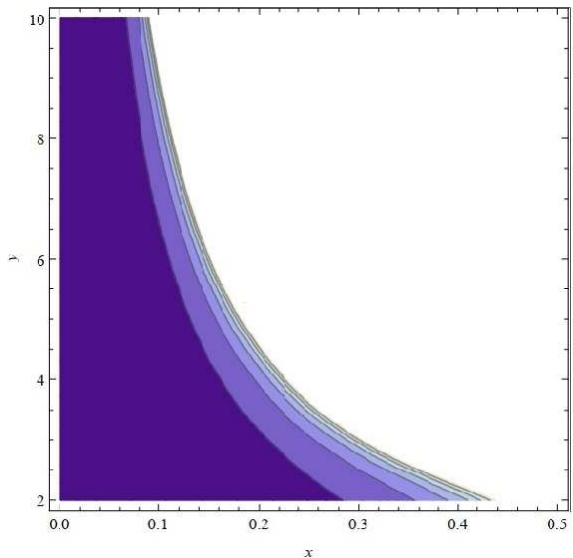}}
  \caption{2D contour plot of $f(x,y)$ for $x\in (0,\frac{1}{2})$ and $y\in (2,\infty )$. The shaded region represents the zone of stability of the thin-shell exclusively under ''mass constraint''.}
\end{figure}

\begin{figure}[!ht]
  \centerline{\includegraphics[scale=0.5]{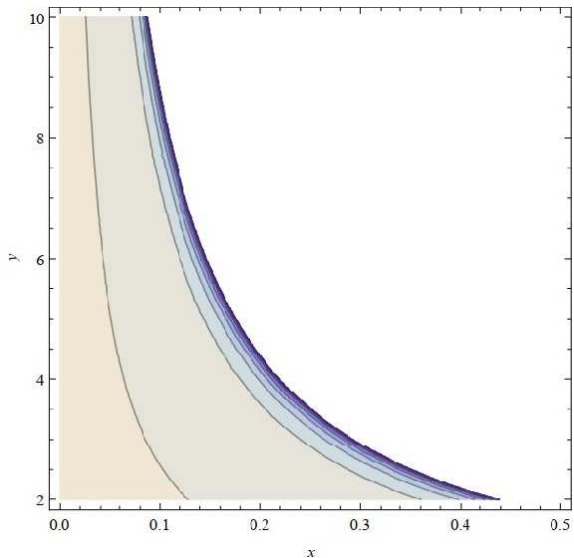}}
  \caption{2D contour plot of $g_{1}(x,y)$. The relevant area is defined by $x\in (\frac{1}{4},\frac{1}{2})$, $y\in (2,4)$ in Case 1 [Sec.3\textit{(a)}]. For Case 2 [Sec.3\textit{(a)}], the relevant area is $x\in (0,\frac{1}{4})$, $y\in (4,\infty )$. In either case, the exclusive "force" constraint indicates that the shaded region is the stable zone.}
\end{figure}

\begin{figure}[!ht]
  \centerline{\includegraphics[scale=0.5]{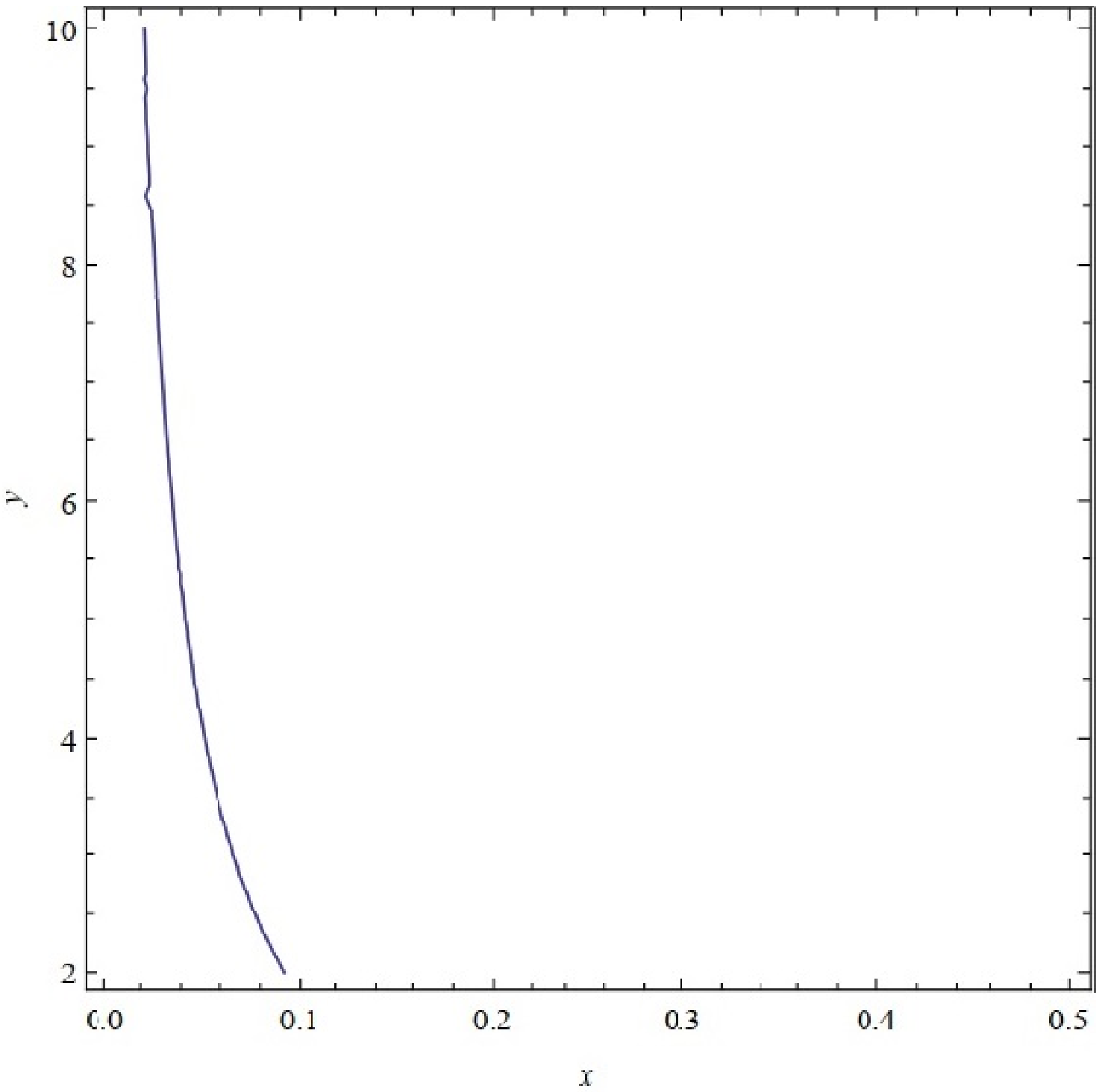}}
  \caption{Thin-shell stability under simultaneous constraints is restricted to parameter values ($x,y$) on the left of the curve obtained by the intersections $f(x,y)\cap g_{1}(x,y)$.}
\end{figure}

It can be noticed from Eq.(26) that for $f(x,y)$ to be real, two conditions
must be fulfilled: one is $x<\frac{1}{2}$, which implies $a_{0}>$ $2M$ as
required, and the other is $y<\frac{1}{x}$ or $a_{0}>r_{0}$. The latter
condition guarantees the reality of also $g_{1}(x,y)$. Note further that $y=%
\frac{r_{0}}{M}$ has no upper bound but has a lower bound $2$ such that the
wormhole throat satisfies $r_{0}>2M$ as required. The above Cases then split
the $xy$ plane into two areas: In Case 1, $x\in (\frac{1}{4},\frac{1}{2})$, $%
y\in (2,4)$ and in Case 2, $x\in (0,\frac{1}{4})$, $y\in (4,\infty )$. In
the contour plots, the two areas are combined such that $x\in (0,\frac{1}{2}%
) $, $y\in (2,\infty )$.

\textit{(b) Schwarzschild black hole-MH2 wormhole }

The second wormhole MH2, cast in the GLV form, is

\begin{equation}
ds^{2}=-\left( \frac{r}{r_{0}}\right) ^{4}dt^{2}+\frac{dr^{2}}{1-\frac{%
r_{0}^{2}}{r^{2}}}+r^{2}(d\theta ^{2}+\sin ^{2}{\theta }d\phi ^{2}),
\end{equation}

\begin{equation}
b_{-}=\frac{r_{0}^{2}}{r},\qquad \Phi _{-}=\frac{1}{2}\ln \left[ \frac{r^{6}%
}{r_{0}^{4}(r^{2}-r_{0}^{2})}\right] .
\end{equation}%
Like in\textit{\ (a)}, stability zone of the thin-shell depends on the sign
of $\Phi ^{\prime }(a_{0})$, which for the MH2 solution is given by%
\begin{equation}
\Phi ^{\prime }(a_{0})=\frac{2a_{0}^{2}-3r_{0}^{2}}{\left(
a_{0}^{2}-r_{0}^{2}\right) a_{0}}.
\end{equation}%
There again occurs two cases. Case 1: $r=a_{0}\in (r_{0},\sqrt{\frac{3}{2}}%
r_{0}]$, for which we get $\Phi ^{\prime }(a_{0})\leq 0$ and Case 2: $%
r=a_{0}\in \lbrack \sqrt{\frac{3}{2}}r_{0},\infty )$, for which we get $\Phi
^{\prime }(a_{0})\geq 0$. In either case, there will occur the effect of
external "force"\ influencing the thin shell motion. These Cases, like in
\textit{(a)}, will split the $xy$ plane into two areas: In Case 1, $x\in (%
\frac{1}{\sqrt{6}},\frac{1}{2})$, $y\in (2,\sqrt{6})$ and in Case 2, $x\in
(0,\frac{1}{\sqrt{6}})$, $y\in (\sqrt{6},\infty )$. In the contour plots to
be given below, the two cases are combined such that $x\in (0,\frac{1}{2})$,
$y\in (2,\infty )$. Introducing the above metric functions in the
inequalities (19,20), we find

\begin{equation}
m_{s}^{\prime \prime }(a_{0})\geq f(x,y)=x\left( \frac{2}{\sqrt{1-2x}}+\frac{%
x}{(1-2x)^{3/2}}+\frac{xy^{2}(2x^{2}y^{2}-3)}{(1-x^{2}y^{2})^{3/2}}\right) ,
\end{equation}

\begin{eqnarray}
\lbrack 4\pi \Xi (a)a]^{\prime \prime }|_{a_{0}} &\geq &g_{2}(x,y)=\frac{%
4-34x^{2}y^{2}+45x^{4}y^{4}-18x^{6}y^{6}}{(1-x^{2}y^{2})^{5/2}}\text{ ,} \\
&&\left( \Phi _{\pm }^{\prime }(a_{0})\leq 0,\text{ Case 1}\right) \\
\lbrack 4\pi \Xi (a)a]^{\prime \prime }|_{a_{0}} &\leq &g_{2}(x,y),\text{ \
\ }\left( \Phi _{\pm }^{\prime }(a_{0})\geq 0,\text{Case 2}\right)
\end{eqnarray}

\begin{figure}[!ht]
  \centerline{\includegraphics[scale=0.5]{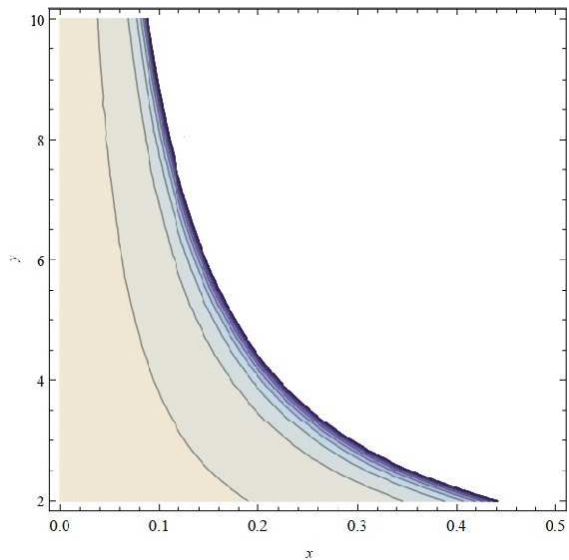}}
  \caption{2D contour plot of $g_{2}(x,y)$. Qualitatively, the stability picture is the same as in Fig.2. For Case 1 [Sec.3\textit{(b)}], the area of $g_{2}(x,y)$ is demarcated by $x\in (\frac{1}{\protect\sqrt{6}},\frac{1}{2})$, $y\in (2,\protect\sqrt{6})$ and for Case 2 [Sec.3\textit{(b)}], the relevant area is $x\in (0,\frac{1}{\protect\sqrt{6}})$, $y\in (\protect\sqrt{6},\infty )$. The "force" constraint implies that the shaded area is the stable zone.}
  \label{Veff}
\end{figure}

\begin{figure}[!ht]
  \centerline{\includegraphics[scale=0.5]{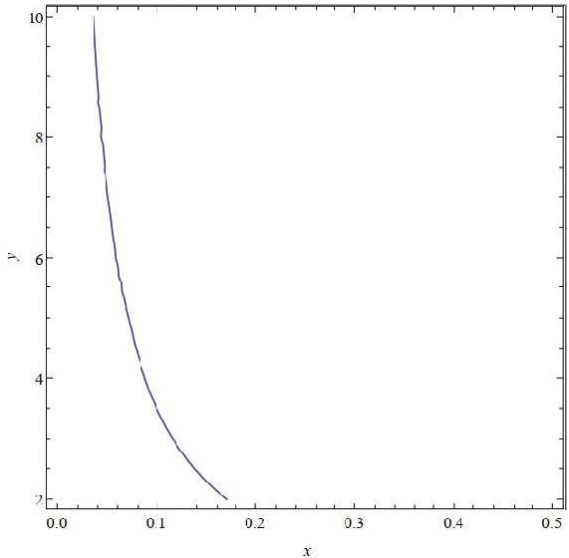}}
  \caption{Thin-shell stability under simultaneous constraints is restricted to parameter values $(x,y)$ to the left of the curve obtained by the intersections $f(x,y)\cap g_{2}(x,y)$.}
\end{figure}

It can be seen that $f(x,y)$ in both \textit{(a)} and \textit{(b)} is the
same, so exactly the same mass constraint applies to both. However, the
functions $g_{1}(x,y)$ and $g_{2}(x,y)$ differ so that the force constraints
different. As a result, the combined effect of two constraints would also
differ leading to different zones of stability. The results are discussed
below.

\section{Results and discussion}

The goal of this work was to demarcate the stability regions of the
thin-shell formed by gluing the Schwarzschild black hole with $f(R)$-gravity
wormholes derived by MH. The remarkable feature of analytical MH wormholes
is that they do not require elusive exotic matter but unfortunately are
unstable. A limited stability ("graceful exit") can however be explored
within the GLV formalism, which is robust providing an excellent way to
tackle wider classes of thin-shell wormholes surgically born out of two
static spherically symmetric spacetimes. The present study yields possible
parameter ranges of the participating solutions (Schwarzschild -MH) for
which stability in some sense can be achieved. Earlier, stability of
thin-shell from Schwarzschild-Schwarzschild gluing using the approach of a
potential was studied before by Poisson and Visser [43]. The stability
constraints on the classes of thin-shells investigated here have relevance
for their very existence, since stable \textit{analytic} wormholes are
rather scarce. The study of the properties of light rings on either side of
the shell would be interesting as a future task in connection with their
characteristic ring-down modes [3-6].

The thin-shell examples considered above are interesting in their own right
since by construction the exterior has a Schwarzschild vaccum ($M\neq 0$).
Physically, this thin-shell configuration resembles a gravastar having an
interior $f(R)$-MH spacetime but the exterior vacuum is indistinguishable
from the Schwarzschild black hole. The stability of the thin-shell is
dictated by two constraints, one is "mass" constraint symbolized by an
explicit inequality involving the second derivative of the mass of the
throat, $m_{s}^{\prime \prime }(a_{0})$ and the other is a external "force"
constraint through the inequalities on $[4\pi \,\Xi (a)\,a]^{\prime \prime }$
depending on the signs of $\Phi _{\pm }^{\prime }(a_{0})$. Note that the
"mass" constraint (19) does not involve $\Phi _{\pm }$ and its derivatives,
while the "force" constraint (20 and its reverse) exists due to $\Phi _{\pm
}\neq 0$. The latter constraint is a new discovery of GLV. Using these two
constraints, we had demarcated the zones of stability in the parameter space.

The analyses in this paper bring out some important characteristics of
Schwarzschild-MH thin-shell wormholes. Even though the two independent MH
wormholes have different metrics, the stability picture of thin-shells under
discussion is remarkably similar. Especially, the contour for $f(x,y)$ is
the same for both thin-shells since $b_{\pm }$ are the same for both MH
solutions, only the $\Phi _{\pm }$ are different. The variables for the
contour plots are chosen as $x=\frac{M}{a_{0}},$ $y=\frac{r_{0}}{M},$ where $%
x\in (0,\frac{1}{2})$, $y\in (2,\infty )$, the ranges being dictated by the
requirement that the gluing radius $a_{0}>2M$ and that the wormhole throat $%
r_{0}>2M$. The reality requirements of the functions $f(x,y)$, $g_{1}(x,y)$
and $g_{2}(x,y)$ yield parameter ranges $x<\frac{1}{2}$ and $y<\frac{1}{x}$.
It is clear that as $x$ increases from $0$ to $\frac{1}{2}$, the gluing
radius $a_{0}$ shrinks from $\infty $ to $2M$. Similarly, when $y$ increases
indefinitely from $2$, the wormhole throat radius $r_{0}$ increases from $2M$
to $\infty $.

The following are our new results. The shaded regions in the 2D contour
plots represent the "topography" of the surfaces $f(x,y)$, $g_{1}(x,y)$ and $%
g_{2}(x,y)$. The higher values on the surfaces are marked by layers of
lighter regions. The zones of stability depend on the Cases discussed in
Sec.3. The "mass" constraint however does not involve such Cases as it does
not involve $\Phi $. Now consider \textit{(a)} of Sec.3 analyzing
Schwarzschild-MH1 thin-shell. The contour plot of the function $f(x,y)$ is
given in Fig.1 for the parameter values $x\in (0,\frac{1}{2})$ and $y\in
(2,\infty )$. The condition (26) then implies that the shaded region in
Fig.1 is the zone of stability. The same conclusion holds for \textit{(b)}
too, i.e., for Schwarzschild-MH2 thin-shell.

The contour plot of the surface $g_{1}(x,y)$ is given in Fig.2, where the
"force" constraint is divided also into two cases as pointed out in Sec.3%
\textit{(a)}. For Case 1, the area is demarcated by $x\in (\frac{1}{4},\frac{%
1}{2})$, $y\in (2,4)$ and for Case 2, area is defined by $x\in (0,\frac{1}{4}%
)$, $y\in (4,\infty )$. (Note that the lines inside plots make boundaries of
points of same heights and not the boundaries of the areas mentioned here).
In either case, the shaded region in Fig.2 indicates the stabilty zone
purely due to "force" constraint. The contour plot of the surface $%
g_{2}(x,y) $ is given in Fig.4 and again there occurs two Cases of
inequalities as pointed out in Sec.3\textit{(b)}. Qualitatively, the
stability picture is similar. For Case 1, the area is demarcated by $x\in (%
\frac{1}{\sqrt{6}},\frac{1}{2})$, $y\in (2,\sqrt{6})$ and for Case 2, area
is $x\in (0,\frac{1}{\sqrt{6}})$, $y\in (\sqrt{6},\infty )$ implying that
the shaded area in Fig.4 is the stable zone. However, the individual
constraints do not completely describe the zones of stability.

The complete stability zone should be determined by the\textit{\ combined}
application of "mass" \ and "force" constraints. In this case, the stability
zone is restricted to parameter values $\left( x,y\right) $ on the left of
the curves obtained by the intersections $f(x,y)\cap g_{1}(x,y)$ (Fig.3) and
$f(x,y)\cap g_{2}(x,y)$ (Fig.5) respectively. The interesting result that
follows from it is that Schwarzschild-MH2 thin-shell has a larger stability
zone (Fig.5) than that of Schwarzschild-MH1 (Fig.3). The intriguing result
that emerges is that the thin-shell motion is stable for $x<0.1$ (Fig.3) and
for $x<0.2$ (Fig.5), i.e., when the gluing radius $a_{0}$ is rather far away
from the horizon $a_{0}=2M$ or $x=\frac{1}{2}$. It can be verified that all
the above stability zones can be read off from the 3D plots as well. Such a
complete stability picture is possible essentially due to the effect of the
"external force" constraint, a new discovery by GLV.

\section*{Acknowledgments}

The reported study was funded by RFBR according to the research project No. 18-32-00377.


\begin{thebibliography}{99}

\bibitem{Mazharimousavi:2016}
  S.H. Mazharimousavi and M. Halilsoy, Mod. Phys. Lett. A \textbf{31}, 1650192 (2016).

\bibitem{Morris:1988}
  M.S. Morris and K.S. Thorne, Am. J. Phys. \textbf{56}, 395 (1988).

\bibitem{Cardoso:2016}
  V. Cardoso, E. Franzin and P. Pani, Phys. Rev. Lett. \textbf{116}, 171101 (2016); \textbf{117}, 089902E (2016).

\bibitem{Cardoso:2017}
  V. Cardoso and P. Pani, Nature Astronomy \textbf{1}, 586 (2017).

\bibitem{Konoplya:2016}
  R.A. Konoplya and A. Zhidenko, JCAP 12 (2016) 043.

\bibitem{Nandi:2017}
  K.K. Nandi, R. N. Izmailov, A.A. Yanbekov and A.A. Shayakhmetov, Phys. Rev. D \textbf{95}, 104011 (2017).

\bibitem{Hochberg:1998}
  D. Hochberg and M. Visser, Phys. Rev. Lett. \textbf{81}, 746 (1998).

\bibitem{Nojiri:2006}
  S. Nojiri and S.D. Odintsov, Phys. Rev. D \textbf{74}, 086005 (2006).

\bibitem{Mazharimousavi:2011}
  S.H. Mazharimousavi and M. Halilsoy, Phys. Rev. D \textbf{84}, 0640328 (2011);

  A. Sheykhi, Phys. Rev. D \textbf{86}, 024013 (2012);

  M. Cvetic, S. Nojiri and S.D. Odintsov, Nucl. Phys. B \textbf{628}, 295 (2002);

  R.G. Cai, Phys. Rev. D \textbf{65}, 084014 (2002).

\bibitem{Cruz:2006}
  A. de la Cruz-Dombriz and A. Dobado, Phys. Rev. D \textbf{74}, 087501 (2006);

  A. de la Cruz-Dombriz, A. Dobado and A.L. Maroto, Phys. Rev. D \textbf{80}, 124011 (2009); Phys. Rev. D \textbf{83}, 029903E (2011);

  A. de la Cruz-Dombriz, A. Dobado and A.L. Maroto, Phys. Rev. Lett. \textbf{103}, 179001 (2009);

  P.K.S. Dunsby, V.C. Busti, S. Kandhai, Phys. Rev. D \textbf{89}, 064029 (2014);

  J.A.R. Cembranos, A. de la Cruz-Dombriz and P. Jimeno Romero, Int. J. Geom. Meth. Mod. Phys. \textbf{11}, 1450001 (2014);

  A. de la Cruz-Dombriz, P.K.S. Dunsby, S. Kandhai, D. S\'{a}ez-G\'{o}mez, Phys. Rev. D \textbf{93}, 084016 (2016).

\bibitem{Olmo:2011}
  G.J. Olmo and D. Rubiera-Garcia, Phys. Rev. D \textbf{84}, 124059 (2011);

  D. Bazeia, L. Losano, G.J. Olmo and D. Rubiera-Garcia, Phys. Rev. D \textbf{90}, 044011 (2014);

  E. Barrientos, F.S.N. Lobo, S. Mendoza, G.J. Olmo and D. Rubiera-Garcia, Phys. Rev. D \textbf{97}, 104041 (2018);

  D. Bazeia, L. Losano, R. Menezes , G.J. Olmo and D. Rubiera-Garcia, Eur. Phys. J. C \textbf{75}, 569 (2015).

\bibitem{Goswami:2014}
  R. Goswami, A.M. Nzioki, S.D. Maharaj and S.G. Ghosh, Phys. Rev. D \textbf{90}, 084011 (2014);

  A.M. Nzioki, R. Goswami and P.K.S. Dunsby, Int. J. Mod. Phys. D \textbf{26}, 1750048 (2016);

  A.M. Nzioki, S. Carloni, R. Goswami and P.K.S. Dunsby, Phys.Rev. D \textbf{81}, 084028 (2010);

  A.M. Nzioki, P.K.S. Dunsby, R. Goswami and S. Carloni, Phys. Rev. D \textbf{83}, 024030 (2011);

  T. Clifton, P. Dunsby, R. Goswami and A.M. Nzioki, Phys. Rev. D \textbf{87}, 063517 (2013).

\bibitem{Sotiriou:2010}
  T.P. Sotiriou and V. Faraoni, Rev. Mod. Phys. \textbf{82}, 451 (2010).

\bibitem{Nojiri:2011}
  S. Nojiri and S.D. Odintsov, Phys. Rept. \textbf{505}, 59 (2011).

\bibitem{Capozziello:2011}
  S. Capozziello and V. Faraoni, \ \textit{Beyond Einstein gravity}, Fundamental Theories of Physics 170 (Springer, 2011).

\bibitem{Sebastiani:2011}
  L. Sebastiani and S. Zerbini, Eur. Phys. J. C \textbf{71}, 1591 (2011).

\bibitem{Bergliaffa:2011}
  S.E.P. Bergliaffa and Y.E.C. de O. Nunes, Phys. Rev. D \textbf{84}, 084006 (2011).

\bibitem{Mazharimousavi:2012}
  S.H. Mazharimousavi and M. Halilsoy, Phys. Rev. D \textbf{86}, 088501 (2012).

\bibitem{Mazharimousavi:2013}
  S.H. Mazharimousavi, M. Kerachian and M. Halilsoy, Int. J. Mod. Phys. D \textbf{22}, 1350057 (2013).

\bibitem{Myung:2011}
  Y.S. Myung, T. Moon and E.J. Son, Phys. Rev. D \textbf{83}, 124009 (2011).

\bibitem{Dolgov:2003}
  A.D. Dolgov and M. Kawasaki, Phys. Lett. B \textbf{573}, 1 (2003).

\bibitem{Olmo:2005}
  G.J. Olmo, Phys. Rev. Lett. \textbf{95}, 261102 (2005).

\bibitem{Faraoni:2005}
  V. Faraoni and S. Nadeau Phys. Rev. D \textbf{72}, 124005 (2005).

\bibitem{Olmo:2007}
  G.J. Olmo, Phys. Rev. D \textbf{72}, 083505 (2005); D \textbf{75}, 023511 (2007).

\bibitem{Myung:2011}
  Y.S. Myung, Eur. Phys. J. C \textbf{71}, 1550 (2011).

\bibitem{Myung:2013}
  Y.S. Myung, Phys. Rev. D \textbf{88}, 104017 (2013).

\bibitem{Nzioki:2014}
  A.M. Nzioki, R. Goswami and P.K.S. Dunsby, Phys. Rev. D \textbf{89}, 064050 (2014).

\bibitem{Mazharimousavi:2011}
  S.H. Mazharimousavi and M. Halilsoy, Phys. Rev. D \textbf{84}, 064032 (2011).

\bibitem{Mazharimousavi:2012}
  S.H. Mazharimousavi, M. Halilsoy and T. Tahamtan, Eur. Phys. J. C. \textbf{72}, 1851 (2012).

\bibitem{Soroushfar:2016}
  S. Soroushfar, R. Saffari and N. Kamvar, Eur. Phys. J. C \textbf{76}, 476 (2016).

\bibitem{Yokokura:2012}
  Y. Yokokura, Int. J. Mod. Phys. A \textbf{27}, 1250160 (2012).

\bibitem{Lobo:2009}
  F.S.N. Lobo and M.A. Oliveira, Phys. Rev. D \textbf{80}, 104012 (2009).

\bibitem{Bhattacharya:2017}
  S. Bhattacharya and S. Chakraborty, Eur. Phys. J. C \textbf{77}, 558 (2017).

\bibitem{Bambi:2016}
  C. Bambi , A. Cardenas-Avendano, G.J. Olmo and D. Rubiera-Garcia, Phys. Rev. D \textbf{93}, 064016 (2016).

\bibitem{Mazharimousavi:2016}
  S.H. Mazharimousavi and M. Halilsoy, Mod. Phys. Lett. A \textbf{31}, 1650203 (2016).

\bibitem{Lobo:2005}
  F. S. N. Lobo and P. Crawford, Class. Quant. Grav. \textbf{22}, 4869 (2005);

  E.F. Eiroa, M.G. Richarte and C. Simeone, Phys. Lett. A \textbf{373}, 1 (2008);

  E.F. Eiroa and G.F. Aguirre, Eur. Phys. J. C \textbf{76}, 132 (2016); Phys. Rev. D \textbf{94}, 044016 (2016).

\bibitem{Eiroa:2005}
  E. F. Eiroa and C. Simeone, Phys. Rev. D \textbf{70}, 044008 (2004); Phys. Rev. D \textbf{71}, 127501 (2005);

  F. Rahaman, M. Kalam and S. Chakraborti, Int. J. Mod. Phys. D \textbf{16}, 1669 (2007);

  Z. Amirabi, M. Halilsoy, S.H. Mazharimousavi, Mod. Phys. Lett. A \textbf{33}, 1850049 (2018);

  S.D. Forghani, S.H. Mazharimousavi and M. Halilsoy, Eur. Phys. J. C \textbf{78}, 469 (2018);

  S.H. Mazharimousavi, M. Halilsoy and S.N.H. Amen. Int. J. Mod. Phys. D \textbf{26}, 1750158 (2017);

  S.H. Mazharimousavi and M. Halilsoy, Int. J. Mod. Phys. D \textbf{27}, 1850028 (2017);

  A. Eid, Indian J. Phys. \textbf{92}, 1065 (2018).

\bibitem{Khaybullina:2014}
  A.R. Khaybullina, G.F. Akhtaryanova, R.F. Mingazova, D. Saha and R.N. Izmailov, Int. J. Theor. Phys. \textbf{53}, 1590 (2014);

  J.P.S. Lemos and F.S.N. Lobo, Phys. Rev. D \textbf{78}, 044030 (2008).

\bibitem{Bronnikov:2010}
  K.A. Bronnikov, M.V. Skvortsova and A.A. Starobinsky, Grav. Cosmol. \textbf{16}, 216 (2010).

\bibitem{Garcia:2012}
  N.M. Garcia, F.S.N. Lobo and M. Visser, Phys. Rev. D \textbf{86}, 044026 (2012).

\bibitem{Deruelle:2008}
  N. Deruelle, M. Sasaki and Y. Sendouda, Prog. Theor. Phys. \textbf{119}, 237 (2008).

\bibitem{Visser:1995}
  M. Visser, \textit{Lorentzian Wormholes-From Einstein To Hawking} (AIP, New York, 1995).

\bibitem{Poisson:1995}
  E. Poisson and M. Visser, Phys. Rev. D \textbf{52}, 7318 (1995).
\end{thebibliography}
\end{document}